\documentclass[12pt]{article}
\usepackage{amssymb}
\usepackage{epsfig}
\voffset= - 0.3in
\hoffset= - 0.6in         

\catcode`\@=11
\def\marginnote#1{}

\newcount\hour
\newcount\minute
\newtoks\amorpm
\hour=\time\divide\hour by60
\minute=\time{\multiply\hour by60 \global\advance\minute by-\hour}
\edef\standardtime{{\ifnum\hour<12 \global\amorpm={am}%
        \else\global\amorpm={pm}\advance\hour by-12 \fi
        \ifnum\hour=0 \hour=12 \fi
        \number\hour:\ifnum\minute<10 0\fi\number\minute\the\amorpm}}
\edef\militarytime{\number\hour:\ifnum\minute<10 0\fi\number\minute}

\def\draftlabel#1{{\@bsphack\if@filesw {\let\thepage\relax
   \xdef\@gtempa{\write\@auxout{\string
      \newlabel{#1}{{\@currentlabel}{\thepage}}}}}\@gtempa
   \if@nobreak \ifvmode\nobreak\fi\fi\fi\@esphack}
        \gdef\@eqnlabel{#1}}
\def\@eqnlabel{}
\def\@vacuum{}
\def\draftmarginnote#1{\marginpar{\raggedright\scriptsize\tt#1}}

\def\draft{\oddsidemargin -.5truein
        \def\@oddfoot{\sl preliminary draft \hfil
        \rm\thepage\hfil\sl\today\quad\militarytime}
        \let\@evenfoot\@oddfoot \overfullrule 3pt
        \let\label=\draftlabel
        \let\marginnote=\draftmarginnote
   \def\@eqnnum{(\theequation)\rlap{\kern\marginparsep\tt\@eqnlabel}%
\global\let\@eqnlabel\@vacuum}  }

\def\appname{Appendix}
\newcounter{app}
\def\theapp{\Alph{app}}
\def\app{\par
   \addvspace{4ex}
   \@afterindentfalse
  \secdef\@app\@dapp}
\def\@app[#1]#2{\ifnum \c@secnumdepth >\m@ne
        \refstepcounter{app}
        \addcontentsline{toc}{app}{\theapp
        \hspace{1em}#1}\else
      \addcontentsline{toc}{app}{ #1}\fi
   {\parindent \z@ \raggedright
    \Large \bf \appname~\theapp .
   \Large  \bf 
    #2}\nobreak
   \vskip 4ex   \noindent
\setcounter{equation}{0}
\def\theequation{\Alph{app}.\arabic{equation}}}
\def\@dapp#1{%
{\parindent \z@ \raggedright  \bf #1}\par\nobreak}
\def\l@app#1#2{\addpenalty{\@secpenalty}%
   \addvspace{1em plus\p@}%
   \begingroup
   \@tempdima 3em
     \parindent \z@ \rightskip \@pnumwidth
     \parfillskip -\@pnumwidth
     { \bf
     \leavevmode
     #1\hfil \hbox to\@pnumwidth{\hss #2}}\par
     \nobreak
   \endgroup}

\makeatletter
\newdimen\normalarrayskip            
\newdimen\minarrayskip               
\normalarrayskip\baselineskip
\minarrayskip\jot
\newif\ifold             \oldtrue            \def\new{\oldfalse}
\def\arraymode{\ifold\relax\else\displaystyle\fi}
\def\eqnumphantom{\phantom{(\theequation)}} 
\def\@arrayskip{\ifold\baselineskip\z@\lineskip\z@
     \else
     \baselineskip\minarrayskip\lineskip1\baselineskip\fi}

\def\@arrayclassz{\ifcase \@lastchclass \@acolampacol \or
\@ampacol \or \or \or \@addamp \or
   \@acolampacol \or \@firstampfalse \@acol \fi
\edef\@preamble{\@preamble
  \ifcase \@chnum
     \hfil$\relax\arraymode\@sharp$\hfil
     \or $\relax\arraymode\@sharp$\hfil
     \or \hfil$\relax\arraymode\@sharp$\fi}}

\def\@array[#1]#2{\setbox\@arstrutbox=\hbox{\vrule
     height\arraystretch \ht\strutbox
     depth\arraystretch \dp\strutbox
width\z@}\@mkpream{#2}\edef\@preamble{\halign \noexpand\@halignto
\bgroup \tabskip\z@ \@arstrut \@preamble \tabskip\z@ \cr}%
\let\@startpbox\@@startpbox \let\@endpbox\@@endpbox
  \if #1t\vtop \else \if#1b\vbox \else \vcenter \fi\fi
  \bgroup \let\par\relax
  \let\@sharp##\let\protect\relax
  \@arrayskip\@preamble}
\def\eqnarray{\stepcounter{equation}%
              \let\@currentlabel=\theequation
              \global\@eqnswtrue
              \global\@eqcnt\z@
              \tabskip\@centering              
              \let\\=\@eqncr
              $$%
            \halign to \displaywidth  \bgroup
             \eqnumphantom \@eqnsel
      \hskip\@centering                               
    $\displaystyle  \tabskip\z@ {##}$%
    &\global\@eqcnt\@ne \hskip 2\arraycolsep
         $ \displaystyle  \arraymode{##}$\hfil
    &\global\@eqcnt\tw@ \hskip 2\arraycolsep
         $\displaystyle\tabskip\z@{##}$\hfil
         \tabskip\@centering
    &{##}\tabskip\z@\cr}
\makeatother

\newfont{\hr}{msbm10}
\newfont{\ams}{msam10}

\font\numbers=cmss12
\font\upright=cmu10 scaled\magstep1
\def\stroke{\vrule height8pt width0.4pt depth-0.1pt}
\def\topfleck{\vrule height8pt width0.5pt depth-5.9pt}
\def\botfleck{\vrule height2pt width0.5pt depth0.1pt}
\def\Zmath{\vcenter{\hbox{\numbers\rlap{\rlap{Z}\kern 0.8pt\topfleck}\kern
2.2pt
                   \rlap Z\kern 6pt\botfleck\kern 1pt}}}
\def\Qmath{\vcenter{\hbox{\upright\rlap{\rlap{Q}\kern
                   3.8pt\stroke}\phantom{Q}}}}
\def\Nmath{\vcenter{\hbox{\upright\rlap{I}\kern 1.7pt N}}}
\def\Cmath{\vcenter{\hbox{\upright\rlap{\rlap{C}\kern
                   3.8pt\stroke}\phantom{C}}}}
\def\Rmath{\vcenter{\hbox{\upright\rlap{I}\kern 1.7pt R}}}
\def\Z{\ifmmode\Zmath\else$\Zmath$\fi}
\def\Q{\ifmmode\Qmath\else$\Qmath$\fi}
\def\N{\ifmmode\Nmath\else$\Nmath$\fi}
\def\C{\ifmmode\Cmath\else$\Cmath$\fi}
\def\R{\ifmmode\Rmath\else$\Rmath$\fi}

\def\d{\partial}

\def\bea{\begin{eqnarray}}
\def\eea{\end{eqnarray}}

\def\beq{\begin{equation}}
\def\eeq{\end{equation}}
\def\ba{\beq\new\begin{array}{c}}
\def\ea{\end{array}\eeq}
\def\be{\ba}
\def\ee{\ea}
\def\F{{\cal F}}
\def\W{{\cal W}}

\def\Q{{\cal Q}}

\def\stackreb#1#2{\mathrel{\mathop{#2}\limits_{#1}}}

\def\res{{\rm res}}

\def\half{{\textstyle{1\over2}}}
\def\ha{{1\over 2}}
\def\N2{${\cal N}=2$}
\def\4N{${\cal N}=4$}
\def\1N{${\cal N}=1$}
\def\1N*{${\cal N}=1^*$}

\def\beq{\begin{equation}}
\def\eeq{\end{equation}}
\def\ba{\beq\new\begin{array}{c}}
\def\ea{\end{array}\eeq}
\def\be{\ba}
\def\ee{\ea}
\def\theequation{\thesection.\arabic{equation}}

\newcommand{\rf}[1]{(\ref{#1})}

\begin{document}


\begin{flushright}
FIAN/TD-05/10\\
ITEP/TH-26/10\\
MPIM 10-123
\end{flushright}
\vspace{1.0 cm}

\renewcommand{\thefootnote}{\fnsymbol{footnote}}
\begin{center}
{\Large\bf On Gauge Theories as Matrix Models\footnote{Based on talks at
{\em Integrable Systems in Quantum Theory }, Leiden, April 2010, {\em Workshop
on Combinatorics of Moduli space}, Moscow, May 2010, {\em Synthesis of Integrabilities
in the Context of Gauge/String Duality}, Moscow, September 2010 and {\em Geometry and Integrable
Systems}, Moscow, December 2010.}
}\\
\vspace{1.0 cm}
{\large A.~Marshakov}\\
\vspace{0.6 cm}
{\em
Theory Department, P.N.Lebedev Physics Institute,\\
Institute of Theoretical and Experimental Physics,\\ Moscow, Russia\\
\medskip
Max Planck Institute for Mathematics,\\
Bonn, Germany
}\\
\vspace{0.3 cm}
{e-mail:\ \ mars@lpi.ru,\ \ mars@itep.ru}
\end{center}
\begin{quotation}
\noindent
The relation between the Seiberg-Witten prepotentials, Nekrasov functions
and matrix models is discussed. The matrix models of Eguchi-Yang type are
derived quasiclassically as describing the instantonic contribution to the
deformed partition functions of supersymmetric gauge
theories. The constructed explicitly exact solution for the case of conformal
four-dimensional theory is studied in detail, and some aspects of its
relation with the recently proposed logarithmic beta-ensembles are considered.
We discuss also the ``quantization'' of this picture in terms of two-dimensional conformal theory with extended symmetry, and stress its difference from common picture of perturbative expansion a la matrix models. Instead, the representation for Nekrasov functions
in terms of conformal blocks or Whittaker vector suggests some nontrivial relation with
Teichm\"uller spaces and quantum integrable systems.
\end{quotation}

\renewcommand{\thefootnote}{\arabic{footnote}}
\setcounter{section}{0}
\setcounter{footnote}{0}
\setcounter{equation}0

\section{Introduction}

The Seiberg-Witten (SW) prepotentials play an important role in studying properties of the
gauge theories at strong coupling \cite{SW,APS} (see also \cite{MY} and references therein
for recent discussion of these issues). Simultaneously, they are beautiful objects
of pure interest for the mathematical physics, and have been recently re-derived
as quasiclassical limit of the Nekrasov instanton partition functions \cite{N,LMN,NO,MN,M,jap,Bra}.
The latter ones have the form of statistical models, where summing is taken over the sets of random partitions or Young diagrams \cite{N,NF}. Originally arisen from the integrals over the instanton moduli spaces in the double-deformed supersymmetric gauge theories, they are related directly to the correlation functions or conformal blocks in two-dimensional
conformal quantum field theories \cite{AGT} (see also discussion of this relation e.g. in \cite{AGT_part}).

Quasiclassics of Nekrasov functions produce the SW geometry, so that the
prepotential appears as a critical value of the free energy functional of the correspondent statistical model \cite{MN,M}. This happens quite similar to the case of quasiclassics in matrix models, though some details - to be discussed below - in these two cases are still quite different.
There have been already many suggestions (see recent attempts e.g. in \cite{Eyn,MM}) to treat the Nekrasov functions in a similar to
the matrix models way beyond the quasiclassical approximation - at least in the sense of perturbative expansion, producing corrections to the prepotential. The relation seems however
to be not very straightforward.

We start the demonstration of this relation in the most old and transparent example.
In fact it becomes more natural, if instead of full
Nekrasov function one would take only its instantonic part, leaving aside the
perturbative contribution. In the simplest case, one gets in such way the Eguchi-Yang (EY)
one-matrix model \cite{EY}, but in the limit of {\em vanishing} size of the matrix \cite{MN,KS}.
The SW periods are introduced in this context in a completely different way,
not being the fractions of condensed eigenvalues, like one used to have for the common matrix models.

Partially this reflects the difference of how the string coupling - parameter of genus or quasiclassical expansion is introduced. For the matrix models it is related to the rank of gauge group,
with sensible expansion only at its large values. In the picture of summation over partitions
the string coupling is rather related to particular combination of the two deformation
parameters, while the rank or matrix size tends to zero. Differently, the same distinction
comes from the completely different role
of the Virasoro constraints in these two theories. For the matrix models the Virasoro constraints are equivalent to the loop equations, giving directly the spectral curve
and an iterative procedure of constructing the perturbative corrections, solving the
loop equations order by order (this can be thought as one of the basic features of the B-type theory). In the case of Nekrasov functions, which rather belong to the A-type string models,
the Virasoro constraints determine only the gravitational dressing of Nekrasov function by
the descendants of unity operator, giving no restriction to the form of this function itself.

We are going to discuss also the relation of SW geometry with the recently proposed logarithmic matrix models \cite{lomamo}. For this purpose we study in detail the theories with the fundamental flavors and present, in particular, the explicit solution for the simplest of them -
an Abelian theory with two flavors. It gives a hint, how the SW geometry can be rewritten in the form close to that of the matrix models, but this does not lead to the full identification of these two pictures. Instead, coming back to the pure gauge theories - where there is even no way for identification of the spectral curves - we show how reformulation of
the SW geometry suggests the way of its natural quantization from the angle of view of the relation with the two-dimensional conformal field theories, and this leads straightforwardly to a natural conjecture for the expression of Nekrasov function in terms of this quantized picture.

\setcounter{equation}0
\section{SW prepotentials as limit of Nekrasov functions}

The instantonic calculus \cite{N} in \N2 supersymmetric gauge theory gives rise \cite{MN,M}
to the extended SW prepotential as a critical value of the
functional $\F \stackreb{\epsilon_{1,2}\to 0}{\sim}\ {1\over \epsilon_1\epsilon_2}\log Z$, having an integral representation
\be
\label{functnl}
\F = {1\over 2}\int dx f''(x)F_{UV}(x) -
{1\over 2}\int_{x_1>x_2} dx_1 dx_2 f''(x_1)f''(x_2) F(x_1-x_2)
\ee
and giving the main contribution to the Nekrasov function $Z(\bullet|\epsilon_1,\epsilon_2)$ at vanishing deformation parameters $\epsilon_{1,2}\to 0$,
when extremized w.r.t. the second derivative of the profile function
$f''(x)={d^2 f\over dx^2}$ of the Young diagram. In \rf{functnl} the bare UV potential equals
\be
\label{fuv}
F_{UV}(x)=\sum_{k>0}t_k{x^{k+1}\over k+1}
\ee
and the kernel
\be
\label{SWkern}
F(x)={x^2\over 2}\left(\log x -{3\over 2}\right)
\ee
comes up from the (generalized) Plancherel measure in the sum over random partitions \cite{NO}.
It is expected, that the main contribution at $\epsilon_{1,2}\to 0$ is
given by some large ``limiting'' Young diagram with the profile $f(x)$,
to be found by solving the variational problem for \rf{functnl} upon normalization conditions imposed as constraints
\be
\label{fa}
a_i = \half\int_{{\bf I}_i} dx\ xf''(x),\ \ \ i=1,\ldots,N
\ee
which can be in standard way taken into account by adding them to the functional with the
Lagrange multipliers
\be
\label{Lagr}
\F \rightarrow \F + \sum_{i=1}^Na^D_i\left(a_i-\ha\int_{{\bf I}_i} dx\ xf''(x)\right)
\ee
(for the case of $U(N)$ gauge theory one has to consider solutions with $N$ cuts
$\{ {\bf I}_i \}$, $i=1\ldots,N$).

The whole
setup for \rf{functnl}-\rf{Lagr} is almost identical \cite{varna} to the standard
quasiclassics of the
matrix models (see e.g. \cite{mig,dv,cm,wdvvbad,KM,mamo}), but with few crucial distinctions:
\begin{itemize}
  \item The Coulomb gas kernel in \rf{functnl} is replaces by a multivalued kernel \rf{SWkern}.
  \item The properties of the double derivative $f''(x)$ of the shape function for the large extremal Young diagram are essentially different from those of the eigenvalue density in matrix models.
\end{itemize}

The extremal equation for the
\rf{functnl} gives the system of $N$ integral equations
\be
\label{exeqN}
\sum_{k>0} t_k z^k-\int dx f''(x)(z-x)\left(\log |z-x|-1\right)
=a^D_i,
\ \ \ \ \ z\in {\bf I}_i,\ \ i=1,\ldots,N
\ee
on each segment of the support. Generally the solution can be expressed in terms of the Abelian integrals on the double cover
\be
\label{dcN}
y^2 = \prod_{i=1}^N(z-x^+_i)(z-x^-_i)
\ee
which is a hyperelliptic curve of genus $g=N-1$.
Define
\be\label{Sfun}
S (z) = F_{UV}'( z)- \int \ dx f''(x) ( z - x) (\log(z - x) - 1) - a^D =
\\
\stackreb{z\to\infty}{=}\ \sum_{k>0} t_k z^k -2N\cdot z(\log z -1) + 2\sum_{i=1}^N a_i\cdot\log z
+ \ldots
\ee
where the integral is taken over the whole support ${\bf I}=\cup_{i=1}^N{\bf I}_i$,
$a^D={1\over N}\sum_{j=1}^N a^D_j$,
and consider its differential, or
\be
\label{dS}
\Phi(z) = {dS\over dz} = \sum_{k>0}kt_k z^{k-1}-\int dx f''(x)\log(z-x)
\ee
satisfying
\be
\label{ficut}
\Phi(x+i0)+\Phi(x-i0)=0,
\ \ \ \ x\in {\bf I}_i,\ \ \ i=1,\ldots,N
\ee
on each cut, and normalized in the following way
\be
\label{ImfiN}
\Phi(x^+_N)=0,\\
\Phi(x^-_j \pm i0) = \Phi(x^+_{j-1} \pm i0) = \pm 2\pi i (N-j+1),\ \ \ \ j=2,\ldots,N
\\
\Phi(x^-_1)=\pm 2\pi i N
\ee
If all $t_k=0$ for $k\neq 1$ and $e^{t_1} = \Lambda^{2N}$, the derivative
$\Phi={dS\over dz}$ is an Abelian integral on the curve
\rf{dcN} with the asymptotic
\be
\label{fiasysw}
\Phi\
\stackreb{P\to P_\pm}{=}\  \mp 2N\log z \pm 2N\log\Lambda  + O(z^{-1})
\ee
($z(P_\pm)=\infty$),
whose jumps are integer-valued due to \rf{ImfiN}, or
$\oint d\Phi \sim 4\pi i\mathbb{Z}$.
It means that the hyperelliptic curve \rf{dcN} can be seen also as an algebraic
Riemann surface for the function
$w=\exp\left(-\Phi/2 \right)$,
satisfying quadratic equation
\be
\label{Todacu}
\Lambda^N\left(w+{1\over w}\right) = P_N(z) = \prod_{i=1}^N (z-v_i)
\ee
or
\be
\label{Today}
y^2 = P_N(z)^2-4\Lambda^{2N}
\ee
i.e. the branch points $\{ x^{\pm}_i \}$ are roots of $P_N(z)\mp 2\Lambda^N=0$, and
\be
\label{yw}
y = \Lambda^N\left(w-{1\over w}\right)
\ee
Note, that in the simplest $N=1$ case the form \rf{Todacu}, i.e.
\be
\label{u1cu}
z-v = \Lambda\left(w+{1\over w}\right)
\ee
can be always achieved by change of the variables, i.e. it can be used as well for the
switched on higher flows \cite{MN,M}. However generally, for nonvanishing higher couplings
in the UV potential \rf{fuv} the profile function cannot be obtained as a jump of
any algebraic function on the curve \rf{dcN}, unlike the case of resolvent and density
in matrix model.

The generating differential \rf{dS} is now
\be
\label{dSsw}
dS = -2\log w dz = -d(2z\log w) + 2z{dw\over w}
\ee
just the Legendre transform
of the SW differential $dS_{SW} \sim z{dw\over w}$
on the curve \rf{Todacu}, \rf{Today}. It periods
\be
\label{SWper}
a_i = {1\over 2\pi i}\oint_{A_i} z {dw\over w}
\ee
coincide with the SW integrals and
the only nontrivial residues at infinity give
\be
\res_{P_+}\left( z^{-1}dS\right) = - \res_{P_-}\left( z^{-1}dS\right) =
\log\Lambda^{2N}
\\
\res_{P_+}\left(dS\right) = - \res_{P_-}\left(dS\right) = 2\sum_{j=1}^N v_j
\ee
The differential \rf{dSsw} satisfies the condition
\be
\delta dS \sim {\delta w\over w} dz = {\delta P(z)\over y} dz\
\stackreb{\sum_{j=1}^N a_j=0}{=}\ {\rm holomorphic}
\ee
where the variation is taken at constant co-ordinate $z$ and constant
scale factor $\Lambda$. This provides the integrability of the gradient
formulas
\be
\label{gradF}
{\d\F\over\d a_i} = \oint_{B_i} z {dw\over w}
\ee
reducing it consistency to the symmetricity of the period matrix - a particular case
of the Riemann bilinear relations.

\setcounter{equation}0
\section{Eguchi-Yang matrix model}

Let us now substitute into \rf{functnl} the shifted profile function \cite{MN,KS}
\be
\label{fg}
f(x) = |x-a|+g(x)
\\
f'(x) = {\rm sign} (x-a) + g'(x) \equiv {\rm sign} (x-a) +
\rho(x)
\\
f''(x) = 2\delta(x-a) + g''(x)
\ee
where the function $g(x)$ and its derivative $g'(x) = \rho(x)$
vanish at the ends of the cut ${\bf I}=(x_-,x_+)$:
\be
g(x_\pm)=0,\ \ \ g'(x_\pm)=\rho(x_\pm)=0
\ee
One obviously gets
\be
\label{functnlg}
\F = F_{UV}(a) + {1\over 2}\int dx g''(x)F_{UV}(x) - \int dx
g''(x)F(x-a) -
\\
- {1\over 2}\int_{x_1>x_2} dx_1 dx_2 g''(x_1)g''(x_2) F(x_1-x_2)
\equiv F_{UV}(a) - \F_{\rm inst}
\ee
where
\be
\label{functnlW}
\F_{\rm inst} = {1\over 2}\int dx \rho(x) \W(x) -
{1\over 2}\int_{x_1>x_2} dx_1 dx_2 \rho(x_1)\rho(x_2) \log(x_1-x_2)
\\
\W(x) = F_{UV}'(x) - 2(x-a)\left(\log(x-a)-1\right)
\ee
The variational problem for \rf{functnlW} is solved under additional
constraint, following from \rf{fa}
\be
\label{g0}
- \int dx\ xg''(x) = \int dx\rho(x) = 0
\ee
The dependence of $\F_{\rm inst}$ upon the variable $a$ can be easily obtained
by redefinition of the $x$-variable $x\to x+a$ and renaming
$\rho(x+a)\to \rho(x)$
\be
\label{Wa}
\F_{\rm inst} = {1\over 2}\int dx \rho(x) \left(F_{UV}'(x+a) - 2x(\log x-1)\right) -
\\
- {1\over 2}\int_{x_1>x_2} dx_1 dx_2 \rho(x_1)\rho(x_2)
\log(x_1-x_2)
\ee
so that
\be
\label{Wda}
{\d\F_{\rm inst}\over \d a} = \half\int dx \rho(x) F_{UV}''(x+a) =
\sum_{k>1}kt_k {\d\F_{\rm inst}\over \d t_{k-1}}
\ee
and it means that
\be
\label{Wt1a}
\F_{\rm inst} (a;t_1,t_2,\ldots) = \left.\widehat{\F}_0 (t_1,t_2,\ldots)\right|_{t_1\to
F_{UV}''(a)}
\ee
and
\be
\widehat{\F}_0 \equiv \left.\F_{\rm inst}\right|_{a=0} = {1\over 2}\int dx \rho(x) \left(F_{UV}'(x)
- 2x(\log x-1)\right) -
\\
- {1\over 2}\int_{x_1>x_2} dx_1 dx_2 \rho(x_1)\rho(x_2)
\log(x_1-x_2)
\ee
is just the effective potential of the EY matrix model
\cite{EY}, constrained by vanishing of the total density of the
eigenvalues \rf{g0}.

It means, that the dependence on
zero Toda time is introduced here in quite uncommon for the matrix
models way \rf{Wda}, \rf{Wt1a}. On the level of the Toda chain hierarchy, which governs
the dynamics over the parameters of the potential \cite{GMMMO}, one deals here with two
completely different solutions to the Toda equation
\be
\label{Todaeq}
{\d^2\F\over\d t_1^2} = \exp\left({\d^2\F\over\d t_0^2}\right)
\ee
namely
\be
\label{fmamo}
\F_{\rm mamo} = \half t_0^2\left(\log t_0-{3\over 2}\right) + \half t_0 t_1^2 + \ldots,\ \ \ \
t_0 = \hbar N
\ee
for the standard matrix model (see e.g. \cite{mamo}), but
\be
\label{fey}
\F_{\rm EY} = \half t_0 t_1^2 + e^{t_1} + \ldots,\ \ \ \ t_0=a
\ee
for the EY model.
The essential difference comes in the zero-time dependence: for the standard matrix model free
energy necessarily contains the function \rf{SWkern} ($F(t_0)$ in the r.h.s. of \rf{fmamo}), giving rise to the logarithm into the second $t_0$-derivative, ``canceled'' further by exponentiating in \rf{Todaeq}, while in the EY case the absence of logarithms of
$t_0=a$ requires necessarily the exponential dependence on $t_1$.
Moreover, the $W$-boson masses, which are
associated to the periods of dual to \rf{Sfun} SW differential, are {\em not} related with the filling fractions of the matrix model with the EY potential.

Note also, since
\be
\W(x) = F_{UV}'(x) + 2a(\log a -1) +
 \sum_{n>0}{x^{n+1}\over n(n+1)a^n}
\ee
the effective potential of the EY matrix model satisfies the property
\be
\label{Wla}
\F_{\rm inst} = \F_{\rm inst}(t_1;{\hat t}_2,{\hat t}_3,\ldots)
\\
{\hat t}_k = t_k + {1\over k(k-1)a^{k-1}},\ \ \ k\geq 2
\ee
and has a natural expansion at large values of $a$.

For the non-Abelian case instead of \rf{fg} one has to make a
substitution
\be
\label{fLg}
f(x) = L(x;{\bf a})+g(x) \equiv \sum_{j=1}^N |x-a_j| + g(x)
\\
f'(x) = \sum_{j=1}^N\ {\rm sign} (x-a_j) + g'(x) \equiv
\sum_{j=1}^N\ {\rm sign} (x-a_j) +
\rho(x)
\\
f''(x) = 2\sum_{j=1}^N\delta(x-a_j) + g''(x)
\ee
after which again the function $g(x)$ and its derivative $g'(x) =
\rho(x)$ vanish at the ends of all cuts ${\bf I}_j=(x^-_j,x^+_j)$, ${\bf I}=
\cup_{j=1}^N {\bf I}_j$:
\be
g(x^\pm_j)=0,\ \ \ g'(x^\pm_j)=\rho(x^\pm_j)=0,\ \ \ j=1,\ldots,N
\ee
Now, instead of \rf{functnlg} one gets
\be
\label{functnLg}
\F = \sum_{j=1}^NF_{UV}(a_j) - \sum_{i\neq j}F(a_i-a_j) +
\\
+{1\over 2}\int dx g''(x)F_{UV}(x) - \int dx g''(x)\sum_{j=1}^N
F(x-a_j) -
\\
- {1\over 2}\int_{x_1>x_2} dx_1 dx_2 g''(x_1)g''(x_2) F(x_1-x_2)
\equiv
\\
= \sum_{j=1}^NF_{UV}(a_j) - \sum_{i\neq j}F(a_i-a_j) - \F_{\rm inst}
\equiv \F_0 - \F_{\rm inst}
\ee
where $\F_0$ is
just the sum of the classical and perturbative contributions to the
prepotential (see Appendix~\ref{ap:pert}), while the instantonic part
is again given by effective potential of a ``matrix model''
\be
\label{functnAW}
\F_{\rm inst} = {1\over 2}\int dx \rho(x) \W(x) -
{1\over 2}\int_{x_1>x_2} dx_1 dx_2 \rho(x_1)\rho(x_2) \log(x_1-x_2)
\\
\W(x) = F_{UV}'(x) - 2\sum_{j=1}^N(x-a_j)\left(\log(x-a_j)-1\right)
\ee
The variational problem for the functional \rf{functnAW} is again solved under
constraint \rf{g0} at each component of the cut ${\bf I}$, since
\be
\label{gaD}
\sum_{j=1}^N a^D_j\left(a_j-\half\int_{{\bf I}_j}dx xf''(x)\right) =
-\half\sum_{j=1}^N a^D_j\int_{{\bf I}_j}dx xg''(x) =
\half\sum_{j=1}^N a^D_j\int_{{\bf I}_j}dx \rho(x)
\ee
We have found therefore, that in the non-Abelian case the instantonic partition function is described by the EY type matrix model with vanishing filling fractions, where
the role of the SW periods is played by the impurities in the potential.

\setcounter{equation}0
\section{Supersymmetric QCD and matrix models}

In order to analyze possible outcome of this similarity, let us now turn to the
case of supersymmetric QCD, or the \N2 supersymmetric Yang-Mills theory with
extra fundamental flavors of matter \cite{HO,GMMM}. We shall concentrate mostly on four-dimensional
conformal theory, i.e. with the number of flavors $N_f=2N$ and the vanishing \N2
beta-function $\beta = 2N-N_f=0$, coming back later to the case of pure supersymmetric gauge theories, after taking the limit of large masses of the fundamental multiplets.

\subsection{Abelian theory with two flavors: explicit solution}

In the case of Abelian theory with two flavors $f=1,2$ the
functional \rf{functnl} is changed just by substitution $F_{UV}(x)\to F_{UV}(x)+\sum_f
F(x-m_f)$, (see Appendix~\ref{ap:pert}, where this is discussed from the point of view of perturbative prepotentials)
\be
\label{functcnf}
\F = {1\over 2}\int dx f''(x)(F_{UV}(x)+\sum_f F(x-m_f)) -
\\
- {1\over 2}\int_{x_1>x_2} dx_1 dx_2 f''(x_1)f''(x_2) F(x_1-x_2) +
a^D\left(a-\half\int dx\ xf''(x)\right)
\ee
so that the function
\be
\label{Sconf}
S(z) = 2{d\over dz}{\delta\F\over\delta f''(z)} = F_{UV}'(z) +
\sum_f F'(z-m_f) - \int F'(z-x)f''(x)dx - a^D =
\\
\stackreb{z\to\infty}{=}\ \sum_{k>0}t_kz^k  +
\left(2a-\sum_f m_f\right)\log z
-\left(a^D +\sum_f m_f\right) + O\left({1\over z}\right)
\\
S(z)\ \stackreb{z\to m_f}{=}\ (z-m_f)\log(z-m_f) +
\ldots,\ \ \ f=1,2
\ee
acquires extra singularities at $z\to m_f$, $f=1,2$, while the
EY $z\log z$-singularity at $z\to\infty$ is canceled due to
vanishing beta-function (for two flavors).

More transparently it is seen for
\be
\label{Phiconf}
\Phi(z) = {dS\over dz} = F_{UV}''(z) +
\sum_f \log(z-m_f) - \int \log(z-x)f''(x)dx =
\\
= \sum_{k>0}kt_kz^{k-1} + \sum_f
\log\left(1-{m_f\over z}\right) - \int \log\left(1-{x\over z}\right)f''(x)dx =
\\
\stackreb{z\to\infty}{=}\ \sum_{k>0}kt_kz^{k-1} + {2a-\sum_f m_f\over
z} + \sum_{k>0}{1\over z^{k+1}}\left(2{\d\F\over\d t_k}-{1\over k}\sum_f m_f^k\right)
\\
\Phi(z)\ \stackreb{z\to m_f}{=}\ \log(z-m_f) + O(1),\ \ \ f=1,2
\ee
and
\be
\label{dPhiconf}
{d\Phi\over dz} = F_{UV}'''(z) +
\sum_f {1\over z-m_f} - \int {f''(x)dx\over z-x} =
\\
\stackreb{z\to\infty}{=}\ \sum_{k>1}k(k-1)t_kz^{k-2} - {2a-\sum_f m_f\over
z^2} + O\left({1\over z^3}\right)
\\
d\Phi(z)\ \stackreb{z\to m_f}{=}\ {dz\over z-m_f} + \ldots,\ \ \
f=1,2
\ee
The solution can be still constructed as an odd under
$y\leftrightarrow -y$ differential on a cylinder
\be
\label{cyl}
y^2=(z-x_+)(z-x_-)
\ee
i.e.
\be
\label{dPhipsi}
d\Phi = {\psi(z)dz\over (z-m_1)(z-m_2)y}
\ee
On small phase space, where only $t_1\neq 0$, the asymptotic
\rf{dPhiconf} requires numerator of \rf{dPhipsi} to be a linear
function
\be
\label{psism}
\left.\psi(z)\right|_{t_k=t_1\delta_{k,1}} = \psi_1z+\psi_0
\ee
where
\be
\label{psi10}
\psi_1=y_1+y_2=\sum_f m_f-2a
\\
\psi_0=-m_1y_2-m_2y_1
\ee
and
\be
\label{yf}
y_f = y(m_f),\ \ \ f=1,2
\ee
The solution \rf{dPhipsi} can be conveniently described by the following
anzatz
\be
\label{anzconf}
d\Phi  = -{1\over\sqrt{1-4\zeta^2}}{dz\over y}{\left(2v-\sum_f
m_f\right)z+2m_1m_2-v\sum_f m_f\over (z-m_1)(z-m_2)} \equiv -
2{dW\over W}
\ee
where we have introduced
\be
\label{WQ}
W+{1\over W} = {z-v\over \sqrt{Q(z)}},
\ \ \ \
Y^2=(1-4\zeta^2)y^2=(z-v)^2-4Q(z)
\\
Q(z) = \zeta^2(z-m_1)(z-m_2),
\ \ \ \
y_f = {m_f-v\over\sqrt{1-4\zeta^2}},\ \ \ f=1,2
\ee
with\footnote{The sign of the root is chosen to fit with the weak-coupling regime, when
$e^{t_1}\to 0$ and $\tanh{t_1\over 2}<0$.}
\be
\label{ztav}
\zeta = {1\over 2\cosh{t_1\over 2}},\ \ \ \ \sqrt{1-4\zeta^2} =
-\tanh{t_1\over 2}
\\
{2v-\sum_f m_f\over\sqrt{1-4\zeta^2}}=-\left(2v-\sum_f m_f\right)\coth{t_1\over 2} = 2a-\sum_f m_f
\ee
At $e^{t_1}\to 0$ the last relation turns into $v=a$, and in this
limit at large masses one can introduce finite scale
$\zeta^2m_1m_2=\Lambda^2$.

In order to write explicitly the generation function \rf{Sconf} one needs to introduce first
the (similar to \rf{u1cu}) uniformization of \rf{cyl} via
\be
\label{wsm}
{L\over 2}\left(\varpi+{1\over \varpi}\right) = z-V,
\ \ \ \
{L\over 2}\left(\varpi-{1\over \varpi}\right) = y,\ \ \ {d\varpi\over \varpi}={dz\over
y}
\\
V = {x_++x_-\over 2}={v-2\zeta^2(m_1+m_2)\over 1-4\zeta^2}
\\
L={x_+ - x_-\over 2} = {2\zeta\over 1-4\zeta^2}\sqrt{\zeta^2(m_1-m_2)^2+(v-m_1)(v-m_2)}
\ee
and define the auxiliary functions
\be
\chi_f={\varpi-\varpi_f\over \varpi\varpi_f-1},\ \ \ \ f=1,2
\\
\chi_f^2-\sigma_f\chi_f+1=0,\ \ \ \sigma_f= {2\over
L}{(m_f-V)(z-V)-L^2\over z-m_f}
\\
\chi_f\ \stackreb{z\to\infty}{=}\ \varpi_f={1\over L}(m_f-V+y_f),\
\ f=1,2
\ee
The generation function \rf{Sconf} now reads
\be
\label{SFm}
S = \sum_{k>0} t_k\Omega_k +(2a-m_1-m_2)\log \varpi - \sum_f(z-m_f)\log\chi_f - y\log\Xi^2
\\
\Omega_k = z^k(\varpi)_+-z^k(\varpi)_-\ ,\ \ \ \ k>0
\\
\zeta={1\over \Xi+{1\over\Xi}},\ \ \ \ \ \log(\varpi_1\varpi_2)={1\over\Xi^2}
\ee
where we have just used the basis of odd under the involution functions on the cylinder \rf{cyl} with the only possible singularities at two infinities, e.g. such as
\be
\Omega_0 = \log \varpi, \ \ \ \Omega_1 = y, \ \ \ \ldots
\ee
If only $t_1\neq 0$, $\Xi^2=e^{t_1}$, and one gets from \rf{SFm}
\be
\label{Sm0}
\left.S(z)\right|_{t_k=t_1\delta_{k,1}} = -2z\log W +(2a-m_1-m_2)\log \varpi +
\sum_{f=1,2} m_f\log\chi_f
\ee
where $W^2=\chi_1\chi_2$ is defined in \rf{WQ}.

Using these formulas it is easy to compute the resulting
prepotential for \rf{anzconf}, \rf{ztav}, which reads (for the only
nonvanishing $t_1$, and up to the linear terms $\sim (m_1+m_2)a$, which do not influence
onto the second derivatives - coupling constants, and can be eliminated by adding
the linear terms $\sim (m_1+m_2)x$ to the potential in \rf{functcnf})
\be
\label{prepm}
\F = \half a^2 t_1 - (a-m_1)(a-m_2)\log\left(1-e^{t_1}\right) +
\F_{\rm pert}(a;{\bf m})
\ee
and contains
\be
\F_{\rm pert}(a;{\bf m}) = \sum_f F(a-m_f)
\ee
so that
\be
\label{Ftau}
\tau(a)={\d^2\F\over\d a^2} = \log{e^{t_1}\prod_f(a-m_f)\over \left(1-e^{t_1}\right)^2}
=\log a^{N_f} + \tau_{\rm conf}
\ee
where
\be
\label{FtauZ}
\tau_{\rm conf} = \log{e^{t_1}\prod_f\left(1-{m_f\over a}\right)\over \left(1-e^{t_1}\right)^2}\
\stackreb{m_f\to 0}{=}\ \log{e^{t_1}\over \left(1-e^{t_1}\right)^2}
\ee
does not depend at vanishing masses on the vacuum expectation values (condensates) and gives rise to
the non-perturbative renormalization of the coupling
\be
\label{Zrentr}
e^{\tau/2} = {1\over e^{-t_1/2}-e^{t_1/2}},\ \ \ {\rm or}
\\
\tau = t_1 - 2\log\left(1-e^{t_1}\right) = t_1 + 2\sum_{k>0}{e^{kt_1}\over k}
\ee
which is a toy-model analog of the instanton renormalization of the coupling constant, given by the Zamolodchikov asymptotic formula \cite{MMMZam}.

The formulas for the EY matrix model \rf{functnlW} in conformal case remain almost
intact, except for the potential
\be
\label{Wm}
\W(x) \rightarrow \W(x;{\bf m}) = F_{UV}'(x) + \sum_f F'(x-m_f) -
2F'(x-a) =
\\
= \sum_{k>0}t_k x^k + \sum_f(x-m_f)\left(\log(x-m_f)-1\right) -
2(x-a)\left(\log(x-a)-1\right)
\ee
For $a=0$ and $m_f=0$, $f=1,2$ the potential \rf{Wm} turns back into
the potential a standard 1-matrix model.

\subsection{Different parameterizations of the curves}

We have already seen in previous section, that to construct the exact solution explicitly one rather needs the
parametrization \rf{wsm}, than more common for supersymmetric QCD form of the curve \rf{WQ}.
Let us point out now, that the curve \rf{WQ}, can be also presented in the form
\be
\label{u1W}
q_1(z)w^2-(z-v)w+q_2(z)=0
\\
q_f(z)=\zeta(z-m_f),\ \ \ f=1,2
\\
w^2 = {q_2\over q_1}W^2 = {q_2\over q_1}\chi_1\chi_2
\ee
Introducing another new variable
\be
\label{xG}
x = {z\over w} = {z\over W}\sqrt{q_1(z)\over q_2(z)}
\ee
one can rewrite \rf{u1W} as
\be
\label{u1G}
x = {\zeta m_1w^2-vw+\zeta m_2\over \zeta w\left(w^2-{w\over \zeta}+1\right)}\
\stackreb{\zeta={1\over \Xi+1/\Xi}}{=}\
{m_1w^2-v\left(\Xi+{1\over\Xi}\right)w+m_2\over w(w-\Xi)\left(w-{1\over\Xi}\right)} =
\\
= {m_2\over w}+{a-m_2\over w-\Xi} - {a-m_1\over w-{1\over\Xi}}
\ee
where we have used \rf{ztav}, (remind also that on small phase space $\Xi^2=e^{t_1}$ is just the exponentiated
UV bare coupling). The SW differential becomes
\be
dS_{SW} = z{dW\over W} = z{dw\over w} + {z\over 2}\left({dq_1\over q_1}-{dq_2\over q_2}\right) =
xdw + {m_1\over 2}{dz\over z-m_1} - {m_2\over 2}{dz\over z-m_2} \equiv
\\ \equiv dS_G + {m_1\over 2}{dz\over z-m_1} - {m_2\over 2}{dz\over z-m_2}
\ee
The first term in the r.h.s. $dS_G = xdw$, and exactly in such form has been introduced in \cite{Gai}, as appearing naturally in the context of D-brane considerations, it is normalized due to \rf{u1G} as
\be
\label{resG}
\res_{w=0} xdw = m_2,\ \ \ \ \res_{w=\infty} xdw = -m_1
\\
\res_{w=\Xi} xdw = a-m_2,\ \ \ \ \res_{w={1\over\Xi}} xdw = -(a-m_1)
\ee
This form of the curve is especially natural from the point of view of comparison with the conformal representation \cite{LMN} in terms of a theory of two-dimensional free scalar field
\be
\label{u1cb}
Z(x;a,m_1,m_2) = Z_{\rm pert}\cdot\langle e^{i m_1\phi(\infty) } e^{i(a-m_1)\phi
(1) } x^{L_0} e^{-i(a-m_2) \phi(1)}e^{ - i m_2\phi(0)} \rangle =
\\
= Z_{\rm pert}\cdot x^{{m_2^2\over 2}+{(a-m_2)^2\over 2}}\langle e^{i m_1\phi(\infty) } e^{i(a-m_1)\phi(1) }e^{-i(a-m_2) \phi(x)}e^{ - i m_2\phi(0)} \rangle =
\\
=Z_{\rm pert}\cdot x^{a^2/2}\left(1-x\right)^{-(a-m_1)(a-m_2)}
\ee
of the $U(1)$ partition function with $N_f=2$. One can absorb the perturbative contribution inside the correlator by redefinition of the scalar product in two-dimensional theory, see the details below.

From \rf{u1G} one also easily finds, that
\be
\label{dGh}
{\d\over\d a} dS_G = {\d x\over\d a}dw = {\left(\Xi-{1\over\Xi}\right)dw\over (w-\Xi)\left(w-{1\over\Xi}\right)} = d\omega
\\
\res_{w=\Xi}\  d\omega = 1,\ \ \ \ \res_{w={1\over\Xi}}\  d\omega = -1
\ee
is just a degenerate analog of the holomorphic differential.

\subsection{Supersymmetric QCD, conformal theory and logarithmic potentials}

In the case of nonabelian supersymmetric QCD (the $N$-cut solution) the formula
for the EY matrix model potential (with fundamental matter) obviously changes for
\be
\label{potconf}
\W(x) = F_{UV}'(x) +
\sum_{f=1}^{N_f} (x-m_f)\left(\log(x-m_f)-1\right)
- 2\sum_{j=1}^N(x-a_j)\left(\log(x-a_j)-1\right)
\ee
It means, that in conformal case $N_f=2N$ the derivative of this
potential does not have any longer a logarithmic singularity at
infinity
\be
\label{Wpr}
W'(x) = F_{UV}''(x) + \sum_f \log(x-m_f) - 2\sum_j\log(x-a_j) =
\\
= F_{UV}''(x) + \sum_f \log\left(1-{m_f\over x}\right) -
2\sum_j\log\left(1-{a_j\over x}\right)
\ee
but still has logarithmic singularities at $x=a_j$ and $x=m_f$.

Generally, for the $U(N)$ theory with $N_f\leq 2N$ flavors on small
phase space one usually starts \cite{HO,GMMM} with the analog of the representation \rf{WQ}
\be
\label{WQN}
W+{1\over W} = {P(z)\over \sqrt{Q(z)}},\ \ \ \ \ Y^2=P(z)^2-4Q(z)
\\
P(z)=z^N - \sum_{k=0}^{N-2} u_k z^k,\ \ \ \
Q(z) = \Lambda^{2N-N_f}\prod_{f=1}^{N_f}(z-m_f)
\ee
with the generating differential
\be
\label{dSqcd}
dS_{SW} = {zdz\over Y}\left(P'(z) - {P(z)Q'(z)\over 2Q(z)}\right)
\ee
In the conformal case $N_f=2N$ instead of the scale parameter $\Lambda$ one gets the
dimensionless constant, and equation \rf{WQN} is changed for
\be
\label{WQc}
W+{1\over W} = {P(z)\over \sqrt{Q(z)}},\ \ \ \ \
Y^2=(1-4\zeta^2)y^2=P(z)^2-4Q(z)
\\
Q(z) = \zeta^2\prod_{f=1}^{2N}(z-m_f)
\ee
For the vanishing masses the differential \rf{dSqcd} becomes holomorphic
\be
\label{dSh}
dS_{SW}\ \stackreb{Q(z) = \zeta z^{N_f}}{=}\ {dz\over Y}\left(zP'(z)
- {N_f\over 2}P(z)\right) =
\\
= {dz\over Y}\left(\left(N - {N_f\over 2}\right)z^N+\sum_{k=0}^{N-2}
\left({N_f\over 2}-k\right)u_k z^k\right)=
\\
\stackreb{N_f=2N}{=}\ {dz\over Y}\sum_{k=0}^{N-2}
\left(N-k\right)u_k z^k
\ee
For example, in the $N=2$, $N_f=4$ case the differential \rf{dSh} is proportional to the only holomorphic differential on torus (see e.g. \cite{MMMZam})
\be
\label{dSh2}
dS_{SW} \sim u{dz\over Y} = {u\over\sqrt{1-4\zeta^2}}{dz\over
y}
\sim {d\xi\over\eta}
\ee
with
\be
\label{zacu}
\eta^2 = \prod_{f=0,1,\lambda}(\xi-q_f) = \xi(\xi-1)(\xi-\lambda)
\\
\zeta^2 = {\lambda\over(1+\lambda)^2} = {1\over\left( \sqrt{\lambda}+{1\over\sqrt{\lambda}}\right)^2}\ \stackreb{\lambda={1\over\sqrt{\Xi}}}{=}\
{1\over \left(\Xi + {1\over\Xi}\right)^2}
\ee
and since $\Xi = e^{\tau_0/2}$ this is identical to the formula \rf{ztav} from the $N=1$ case.

However, for the nonvanishing masses it is more convenient to use the analog of the
representation \rf{u1W}
\be
\label{uNt}
q(z)w^2-P(z)w+{\tilde q}(z)=0
\ee
where $P(z)$ is the same polynomial of power $N$ as in \rf{WQN}, while for $N_f=2N$ one can naturally introduce
\be
\label{qtq}
q(z)=\zeta \prod_{f=1}^{N}(z-m_f),\ \ \ \
{\tilde q}(z)=\zeta \prod_{f=1}^{N}(z-{\tilde m}_f)
\ee
for some natural splitting of the $N_f=2N$ matter multiplets into $N$ ``fundamental'' and
$N$ ``anti-fundamental'' - with the masses $m$ and ${\tilde m}$ correspondingly. Again, like
in \rf{u1W}, we have introduced here
\be
\label{tW}
w^2 = {{\tilde q}(z)\over q(z)}W^2
\ee
For $N=2$ in \rf{qtq}, equation \rf{uNt} becomes as well a quadratic equation in $z$-variable,
and can be therefore written \cite{EM} as
\be
\label{uNz}
C_2(w)z^2-C_1(w)z+C_0(w)=0
\ee
with
\be
\label{C2}
C_2(w) = \zeta w^2 - w + \zeta\ \stackreb{\zeta=\Xi + {1\over\Xi}}{=}\
\left(\Xi + {1\over\Xi}\right)\left(w-\Xi\right)\left(w-{1\over\Xi}\right)
\ee
Equation \rf{uNz} can be further re-written as
\be
\label{uNtz}
{\tilde z}^2 \equiv \left(z-{C_1\over 2C_2}\right)^2 =
{C_1^2\over 4C_2^2}-{C_0\over C_2} \equiv x^2w^2
\ee
or
\be
\label{uNx}
x^2  = {C_1(w)^2\over 4w^2C_2(w)^2}-{C_0(w)\over w^2C_2(w)}
\ee
At vanishing masses the last equation turns into
\be
\label{vama}
x^2 = -{C_0(w)\over w^2C_2(w)} = - {u\over\zeta w(w-\Xi)\left(w-{1\over\Xi}\right)} =
- {u\over\zeta\Xi^3}{1\over\xi(\xi-1)(\xi-\lambda)}
\ee
in the notations of the elliptic curve \rf{zacu}.

Generally, instead of the holomorphic differential \rf{dSh2} one can write
\be
\label{dSG2}
dS_{SW} = {\sqrt{p(\xi)}d\xi\over\xi(\xi-1)(\xi-\lambda)}
\ee
for some polynomial
\be
\label{pG2}
p(\xi)=\sum_{j=0}^4 p_j\xi^j\ \stackreb{m_f\to 0}{\rightarrow}\
C\cdot \xi(\xi-1)(\xi-\lambda)
\ee
reproducing the holomorphic differential \rf{dSh2} in the limit of vanishing masses.
Due to relation between the coefficients of $p(\xi)$ and
the residues
\be
\res_{\xi=0}\ dS_{SW} = \sqrt{p(0)} = \sqrt{p_0} = \lambda m_0
\\
\res_{\xi=\infty}\ dS_{SW} = \sqrt{p_4} = - m_\infty
\\
\res_{\xi=1}\ dS_{SW} = \sqrt{p(1)} = (1-\lambda) m_1
\\
\res_{\xi=\lambda}\ dS_{SW} = \sqrt{p(\lambda)} = \lambda(\lambda-1)
m_\lambda
\ee
one gets
\be
\label{varp}
\delta_{\rm moduli}p(\xi) = \delta_{\rm moduli}C\cdot \xi(\xi-1)(\xi-\lambda)
\ee
and it means that for the variation of the SW differential one can write
\be
\delta_{\rm moduli}dS_{SW} \sim {d\xi\over\sqrt{p(\xi)}}{\delta_{\rm moduli}p(\xi)\over
\xi(\xi-1)(\xi-\lambda)} = \delta_{\rm moduli}C{d\xi\over\sqrt{p(\xi)}}
\ee
so that the r.h.s. in this equality is obviously holomorphic.

One can also present the 2-differential ${\cal
T}(d\xi)^2=(dS_{SW})^2$ as
\be
\label{Tcoma}
{\cal T} = \left({dS_{SW}\over d\xi}\right)^2 = {p(\xi)\over
\xi^2(\xi-1)^2(\xi-\lambda)^2} =
\\
= \sum_{f=0,1,\lambda}\left({m_f^2\over (\xi-q_f)^2} + {C_f\over
\xi-q_f}\right) = \left(\sum_{f=0,1,\lambda}{m_f\over
\xi-q_f}\right)^2 + {{\cal C}(\xi)\over \xi(\xi-1)(\xi-\lambda)}
\ee
where
\be
\label{Cm}
\sum_{f=0,1,\lambda} C_f = 0,
\ \ \ \
C_1 + \lambda C_\lambda = -\sum_{A=0,1,\lambda\infty}m_A^2
\ee
and
\be
{\cal C}(\xi) = -\xi
\left[\left(\sum_{f=0,1,\lambda}m_f\right)^2 + m_\infty^2\right] +
\left[C_0-2m_0m_1\lambda - 2m_0m_\lambda\right]
\ee
is a linear function satisfying $\delta_{\rm moduli}{\cal
C}(\xi)=\lambda\delta_{\rm moduli}C_0$. Formula \rf{Tcoma} may be assigned with
two possible interpretations:
\begin{itemize}
  \item It has a sense of average of the stress-energy tensor in some two-dimensional conformal field theory
\be
\label{Tconf}
{\cal T}(\xi) = {\langle T(\xi) \prod_{A=0,1,\lambda,\infty}
V_A(q_A)\rangle\over\langle \prod_{A=0,1,\lambda,\infty}
V_A(q_A)\rangle}
\ee
with the insertions of four primary operators of dimensions $m_A^2$, $A=0,1,\lambda\infty$ ($m_\infty$ can be determined from the second equation in \rf{Cm}) and $C_f$ being the corresponding accessor parameters (see e.g. \cite{ZamZam} and references therein);
  \item The r.h.s. \rf{Tcoma} has an obvious form of the matrix model curve, if one takes
  formally in the matrix model the following logarithmic potential
\be
{\cal V}(\xi) = \sum_{f=0,1,\lambda}m_f\log(\xi-q_f)
\ee
and the SW periods can be then identified with the filling fractions \cite{lomamo}.
\end{itemize}
We do not find the last observation to be very useful, despite of existing vast list of literature on the subject, and in the rest of the paper we would discuss rather the first one, digressing it back to the case of the pure supersymmetric gauge theory.

\setcounter{equation}0
\section{Pure gauge theories and Whittaker vectors
\label{ss:whit}}

If one takes in \rf{u1W} the decoupling matter limit $\zeta\to 0$ and $m_{1,2}\to\infty$
so that $\Lambda^2 = \zeta^2m_1m_2= {\rm fixed}$~\footnote{Literally, one needs to perform this limit
in symmetric way, putting $\Lambda=-\zeta m_1=-\zeta m_2$.
}, one gets instead of \rf{u1G} the equation
\be
\label{xu1}
x = {\Lambda\over w^2} + {v\over w} + \Lambda =
{\langle\Psi| J(w) |\Psi\rangle\over \langle\Psi|\Psi\rangle } \equiv \langle J(w) \rangle
\ee
which is an average of the $U(1)$-current $J(w) = \sum_{n\in\mathbb{Z}}{J_n\over w^{n+1}}$
over the state, satisfying
\be
\label{psiu1}
J_n| \Psi\rangle = 0,\ \ \ n>1
\\
J_1| \Psi\rangle = \Lambda| \Psi\rangle,\ \ \ J_0| \Psi\rangle = a| \Psi\rangle
\ee
Since $[J_n,J_m]=n\delta_{n+m,0}$, equations \rf{cohe} can be immediately solved explicitly
by
\be
\label{cohe}
| \Psi\rangle = \Lambda^{L_0}e^{J_{-1}}| a\rangle
\\
J_n| a\rangle = 0,\  n>0,\ \ \ \ \ \ \ J_0| a\rangle = a| a\rangle
\ee
i.e. in terms of the coherent state in the (charged) Fock module, so that
\be
\langle\Psi|\Psi\rangle = \langle a|e^{J_1} \Lambda^{2L_0}e^{J_{-1}}|a \rangle
= \Lambda^{a^2}e^{\Lambda^2}\ \stackreb{\Lambda^2=e^{t_1}}{=}\ \exp\left(\half a^2t_1+e^{t_1}\right)
\ee
The generating differential can be treated as just as an average of the
$\widehat{U(1)}$-current
\be
\label{JdS}
\langle J(w) \rangle = dS_{SW} = xdw
\ee
and its more conventional form \rf{u1cu} is restored by substitution $x={z\over w}$.

For the pure $U(2)$ supersymmetric gauge theory instead of \rf{xu1} one gets an equation \cite{Gai}
\be
\label{x2vir}
x^2 = {\Lambda^2\over w^3} + {u\over w^2} + {\Lambda^2\over w} =
{\langle\Psi| T(w) |\Psi\rangle\over \langle\Psi|\Psi\rangle } \equiv \langle T(w) \rangle
\ee
which has an obvious sense of averaging of the stress-tensor
$T(w) = \sum_{n\in\mathbb{Z}}{L_n\over w^{n+2}}$
over the Whittaker \cite{Bra,GWtk} state $| \Psi\rangle \in {\cal H}_{\Delta,c}$ in the Verma module of the Virasoro algebra with the highest weight $\Delta$ and the central charge $c$, satisfying $L_n| \Psi\rangle = 0$ if $n>1$ and
\be
\label{virpsi}
L_1| \Psi\rangle = \Lambda^2| \Psi\rangle
\ee
An important consequence of \rf{x2vir}, proven in \cite{MMMWtk}, is that the
Nekrasov function of the corresponding theory is given by the scalar product, or the matrix element
\be
\label{NG}
Z(a;\epsilon_1,\epsilon_2) =\langle \Psi | \Psi\rangle =
\langle \Psi_1 |\Lambda^{4L_0} | \Psi_1\rangle
\ee
where
\be
\label{psi1}
| \Psi_1\rangle = \left.| \Psi\rangle\right|_{\Lambda=1}
\ee
after identification
\be
\label{idu2}
\Delta = -{a^2\over\epsilon_1\epsilon_2} + {\epsilon^2\over 4\epsilon_1\epsilon_2},
\ \ \ \
c=1+\frac{6\epsilon^2}{\epsilon_1\epsilon_2},\ \ \ \epsilon=\epsilon_1+\epsilon_2
\ee
Moreover
\be
\label{hw}
\left.| \Psi\rangle\right|_{\Lambda=0} = | \Omega\rangle\in {\cal H}_{\Delta,c}
\\
L_n| \Omega\rangle = 0, n>0,\ \ \ \ L_0| \Omega\rangle =\Delta| \Omega\rangle
\ee
i.e. $|\Omega\rangle$ is proportional to the highest-weight vector,
but with a non-standard normalization
\be
\langle\Omega|\Omega\rangle \sim \Gamma_2(a|\epsilon_1,\epsilon_2)\Gamma_2(-a|\epsilon_1,\epsilon_2)
\ee
to the product of the Barnes double-gamma functions \cite{Barnes} (some details and other references
are collected in Appendix).

The nontrivial part of \rf{x2vir} contains the only equality
\be
{\langle\Psi| L_0 |\Psi\rangle\over \langle\Psi|\Psi\rangle } = {1\over 4}{\d\over\d\log\Lambda}\langle\Psi|\Psi\rangle = u\ \stackreb{\epsilon_{1,2}\to 0}{=}\
{1\over 4}{\d\over\d\log\Lambda}\F_{SW}
\ee
being a sort of renormalization group equation. Again, by $x={z\over w}$ the curve \rf{x2vir} turns into that with $N=2$ from the family \rf{Todacu}, and the generating differential is now  \cite{Gai}
\be
\sqrt{\langle T(w) \rangle}dw = dS_{SW} = xdw = z{dw\over w}
\ee
Generally for the pure $U(N)$ supersymmetric gauge theory one should write the average of
a formal differential operator
\be
\label{Doper}
\langle\Psi| {\cal D}_N|\Psi\rangle =0
\\
{\cal D}_N \equiv D^N-T(w)D^{N-2}-{\cal W}^{(3)}(w)D^{N-3}-\ldots-{\cal W}^{(N)}(w)
\ee
with the coefficients acting in the highest-weight module ${\cal H}_{{\bf a},c}$ of the
$W_N$-algebra, and
$|\Psi\rangle$ is now the Whittaker vector
\be
\label{DW}
{\cal W}^{(N)}_1|\Psi\rangle = \Lambda^N|\Psi\rangle
\\
{\cal W}^{(N)}_n|\Psi\rangle = 0,\ n>1,\ \ \ \ {\cal W}^{(K)}_n|\Psi\rangle = 0,\ n>0,\ K<N
\ee
in the module with the highest weight $|\Omega\rangle = \left.|\Psi\rangle\right|_{\Lambda=0}$
\be
{\cal W}^{(K)}_0|\Omega\rangle = s_K({\bf a};\epsilon_1,\epsilon_2)|\Omega\rangle =
\sum_{j=1}^N \left({a_j^K\over\epsilon_1\epsilon_2}+\ldots\right)|\Omega\rangle,\ \ \ K=1,\ldots,N
\ee
where also everywhere
\be
\label{Dx}
D|\Psi\rangle = x|\Psi\rangle
\ee
This set of equations (again by substitution $x={z\over w}$) gives rise to the SW curves \rf{Todacu} endowed with generating differential $dS_{SW} = xdw=z{dw\over w}$, with a natural
conjecture
\be
\label{ziconj}
Z = \langle\Psi|\Lambda^{2NL_0}|\Psi\rangle
\ee
provided by normalization
\be
\langle\Omega|\Omega\rangle = \left.\langle\Psi|\Psi\rangle\right|_{\Lambda=0} \sim \prod_{i,j}\Gamma_2(a_i-a_j|\epsilon_1,\epsilon_2) \sim Z_{\rm pert}
\ee
written again in terms of the Barnes double-gamma function.

In the $N=1$ and $N=2$ cases equations \rf{DW} (i.e. $\langle\Psi| D-J(w)|\Psi\rangle =0$ for
$W_1=\widehat{U(1)}$ and
$\langle\Psi| D^2-T(w)|\Psi\rangle =0$ for $W_2=Vir$ can be explicitly solved w.r.t. $x =\langle D\rangle$ variable, and we come back to \rf{xu1} and \rf{x2vir} correspondingly. To switch on higher times one just has to generalize the Hamiltonian in
\rf{ziconj}
\be
\label{ztconj}
Z(t_1,\ldots,t_N) \sim \langle\Psi|\exp\left(\sum_{K=1}^N t_K{\cal W}^{(K)}_0\right)|\Psi\rangle
\ee
by introducing higher $W$-flows. Arising of the Gelfand-Dikij-type operator \rf{Doper} in this context together with the $W$-flows in \rf{ztconj} suggests a nontrivial connection with the so-called
$W$-gravity \cite{Wgrav} and geometry of generalized Teichm\"uller spaces.

\section{Conclusion}

We have discussed in these notes some relations between the formulation of supersymmetric gauge theories, coming from the instanton partition functions, and the matrix
models. As we pointed out - there are several parallels of this kind, but neither of them
seems to be very essential by itself.

Let us finish instead with the following important remark. The quasiclassical picture
of the matrix model is indeed very similar to the SW theory, and they are both described
in terms of a quasiclassical integrable system, so that the prepotential
is in fact a restricted Krichever tau-function \cite{KriW}. The simplest part of integrable
dynamics in this case is parameterized by the parameters of potential \rf{fuv}, and reproduces in the simplest case the well-known hierarchy of the dispersionless Toda equation \rf{Todaeq} (for
the higher-genera analogs of dispersionless equations see \cite{KMZ}). The dependence on smooth periods \rf{SWper} is far more transcendental
and, on quasiclassical level, the knowledge is basically exhausted by the gradient formulas \rf{gradF}
and their consequences, like residue formulas, the WDVV equations, etc (see e.g. \cite{KriW,wdvvbad,mamo}).

The formulas of sect.~\ref{ss:whit} demonstrate, that the nontrivial integrable dynamics over quasiclassical
period variables has even far less trivial dispersive analogs. While the naive ``quantization'' of
the dynamics over the times \rf{fuv} leads (in absence of the ``smooth variables'' \rf{SWper})
just to the
string solution of the full hierarchy of the Toda or KP type with the tau-functions presented by
the matrix elements in the two-dimensional theory of free fermions, the dispersive analogs of
the full Krichever tau-function are matrix elements in non-trivial two-dimensional conformal theory, generally with extended symmetry. They also seem to be directly related with the quantization of Teichm\"uller spaces \cite{qute} (higher Teichm\"uller spaces for the case of $W$-gravity), Liouville theory \cite{ZamZam,AlZLi}, and quantum-mechanical integrable dynamics in the systems of Toda type (see e.g. \cite{KLST,NS,KT}). All these relations clearly deserve further investigation, and we are going to return to them elsewhere.

\section*{Acknowledgements}

I am grateful to A.~Braverman, L.~Chekhov, T.~Eguchi, V.~Fock, H.~Itoyama, H.~Kanno, S.~Kharchev, I.~Krichever, A.~Mironov and A.~Morozov for the illuminating discussions.

The work was partly supported by Russian Federal Nuclear Energy Agency,
by the Russian Ministry of Education and Science under the
contract 02.740.11.0608, by RFBR grant 10-02-01315, by joint grants
09-02-93105-CNRSL, 09-02-91005-ANF, 10-02-92109-Yaf-a and by the Max Planck Society. I am grateful to the Yukawa Institute for Theoretical Physics in Kyoto and the Max Planck Institute for Mathematics in Bonn, where essential parts of this work have been done, for the warm hospitality.

\section*{Appendix}
\appendix

\setcounter{equation}0
\section{Perturbative prepotentials
\label{ap:pert}}

The instantonic expansion $\F = \sum_{k\geq 0}\F_k$ in the
non-Abelian theory starts with the perturbative contribution (which does not take into account
the topologically nontrivial configurations of the gauge fields)
\be
\label{prepert}
\F_0 = \sum_{j=1}^N F_{UV}(a_j) -
\sum_{i\neq j}F(a_i-a_j)
\ee
defined entirely in terms of the ultraviolet or classical \rf{fuv} and
perturbative \rf{SWkern} prepotentials. It is totally characterized by
degenerate differential of
\rf{dS}
\be
\label{dPhi0}
d\Phi_0 = F_{UV}'''(z)dz - 2{dP_N(z)\over P_N(z)} = F_{UV}'''(z)dz -
2\sum_{j=1}^N{dz\over z-v_j}
\ee
and the coefficients of the
polynomial $P_N(z)$ \rf{dPhi0} (the same as in \rf{Todacu}) coincide with the
perturbative values of the SW periods
\be
\label{avpert}
a_i = - \half\res_{v_i} zd\Phi_0 = v_i,\ \ \ \ i=1,\ldots,N
\ee
The perturbative generating differential is $dS_0=\Phi_0 dz$, with
\be
\label{Phi0}
\Phi_0 = F_{UV}''(z) - 2\sum_{j=1}^N\log\left( z-v_j\right)
\ee
and satisfies
\be
{\d dS_0\over \d a_j} = 2{dz\over z-v_j},\ \ \ \ j=1,\ldots,N
\\
{\d dS_0\over \d t_k} =  kz^{k-1}dz,\ \ \ \  k>0
\ee
what gives rise to
\be
S_0(z) = F_{UV}'(z) - 2\sum_{j=1}^N(z-v_j)\left(\log(z-v_j)-1\right)
\ee
Equations \rf{gradF}
\be
\label{SWpertS}
a^D_j = {\d \F_0\over\d a_j} = S_0(a_j)
\ee
completely determine \rf{prepert}, since on the perturbative stage one makes no
difference between $v_j$ and $a_j$.

For the theory with $N_f$ fundamental multiplets instead of the formula
\rf{prepert} one has
\be
\label{prepem}
\F_0=\F_{\rm cl}+\F_{\rm pert} = \sum_{j=1}^N F_{UV}(a_j) - \sum_{i\neq j}^N F(a_i-a_j) +
\sum_{f=1}^{N_f}\sum_{j=1}^N F(a_j+m_f)
\ee
which can be obtained from \rf{prepert} just by formal modification of the
UV prepotential via
\be
\label{modUV}
F_{UV}(x) \rightarrow F_{UV}(x) + \sum_{f=1}^{N_f} F(x+m_f)
\ee
which can be further used, after its substitution to the functional \rf{functnl}
to compute the full partition function for the theory with matter.

The ``quantum'' - or, better, double-deformed - version of the perturbative prepotentials
\rf{prepert}, \rf{prepem} can be written in terms of the Barnes double-gamma functions
\be
\label{barnes}
\Gamma_2(x|\epsilon_1,\epsilon_2) \sim \prod_{n,m\geq 0}{1\over x+n\epsilon_1+m\epsilon_2}
\ee
where the infinite product can be understand, say, via the zeta-regularization (see e.g. \cite{Spre})
\be
\label{logG2}
\log\Gamma_2(x|\epsilon_1,\epsilon_2) =
\left.{d\over ds}\sum_{n,m\geq 0}\left(x+n\epsilon_1+m\epsilon_2\right)^{-s}\right|_{s=0}
\ee
analytically continued to $s=0$. The relation to \rf{prepert}, \rf{prepem} is established
via the asymptotic
\be
\label{aslogG2}
\log\Gamma_2(x|\epsilon_1,\epsilon_2)\ \stackreb{\epsilon_{1,2}\to 0}{=}\
{1\over\epsilon_1\epsilon_2}F(x)  + {\rm less\ singular} =
\\
= -{1\over\epsilon_1\epsilon_2}{x^2\over 2}\left(\log x-{3\over 4}\right) + \ldots
\ee



\begin{thebibliography}{7799}

\bibitem{SW}
N.~Seiberg and E.~Witten,
Nucl.\ Phys.\ B {\bf 426}, 19 (1994)
[arXiv:hep-th/9407087]; Nucl. Phys. {\bf B431}, 484  (1994)
[hep-th/9408099].

\bibitem{APS}
P.~Argyres, M.~Plesser and N.~Seiberg, Nucl. Phys. {\bf B471} (1996)
159, [arXiv:hep-th/9603042].

\bibitem{MY}
A.~Marshakov and A.~Yung,
  Nucl. Phys. {\bf B 831} (2010) 72-104,
  arXiv:0912.1366 [hep-th];\\
 A.~Marshakov,
  Theor. Math. Phys. {\bf 165} (2010), 488–502, arXiv:1003.2089 [hep-th].

\bibitem{N}
  N.~Nekrasov,
  Adv.\ Theor.\ Math.\ Phys.\  {\bf 7} (2004) 831
  [arXiv:hep-th/0206161].

\bibitem{LMN}
A.~S.~Losev, A.~Marshakov and N.~Nekrasov, in Ian Kogan memorial volume
{\it From fields to strings: circumnavigating theoretical physics},
581-621; hep-th/0302191

\bibitem{NO}
  N.~Nekrasov and A.~Okounkov,
  arXiv:hep-th/0306238.

\bibitem{MN} A.~Marshakov and N.~Nekrasov,
  JHEP {\bf 0701} (2007) 104, hep-th/0612019;\\
  A.~Marshakov,
  Theor.Math.Phys.  {\bf 154} (2008) 362,
  arXiv:0706.2857.

\bibitem{M}
A.~Marshakov,
  JHEP {\bf 0803} (2008) 055,
  arXiv:0712.2802 [hep-th].

\bibitem{Eyn}
B.~Eynard, M.~Marino and N.~Orantin,
  JHEP {\bf 0706} (2007) 058
  [arXiv:hep-th/0702110];\\
B.~Eynard, J.Stat.Mech. {\bf 0910} (2009) P10011, arXiv:0905.0535;\\
L.~Chekhov, B.~Eynard, O.~Marchal, arXiv:1009.6007.

\bibitem{MM}
 A.~Mironov, A.~Morozov and Sh.~Shakirov,
  JHEP {\bf 1002} (2010) 030
  [arXiv:0911.5721 [hep-th]];
  arXiv:1011.5629 [hep-th].


\bibitem{jap}
 T.~Nakatsu and K.~Takasaki,
  Commun.\ Math.\ Phys.\  {\bf 285} (2009) 445
  [arXiv:0710.5339 [hep-th]];\\
T.~Nakatsu, Y.~Noma and K.~Takasaki,
  Int.\ J.\ Mod.\ Phys.\  A {\bf 23} (2008) 2332
  [arXiv:0806.3675 [hep-th]];
  Nucl.\ Phys.\  B {\bf 808} (2009) 411
  [arXiv:0807.0746 [hep-th]].

\bibitem{Bra}
A.~Braverman, arXiv:math/0401409;\\
A.~Braverman and P.~Etingof, arXiv:math/0409441.

\bibitem{NF}
R.~Flume and R.~Pogossian, Int.J.Mod.Phys. {\bf A18} (2003) 2541;\\
H.~Nakajima and K.~Yoshioka, math/0306198, math/0311058;\\
U.~Bruzzo and F.~Fucito, Nucl.Phys. {\bf B678} (2004) 638-655, math-ph/0310036;\\
F.~Fucito, J.~Morales and R.~Pogossian, JHEP, {\bf 10} (2004) 037, hep-th/040890;\\
S.~Shadchin, SIGMA {\bf 2} (2006) 008, hep-th/0601167

\bibitem{AGT}
L.~Alday, D.~Gaiotto and Y.~Tachikawa,
Lett. Math. Phys. {\bf 91} (2010) 167-197, arXiv:0906.3219

\bibitem{AGT_part}
N.~Wyllard,
JHEP {\bf 0911} (2009) 002, arXiv:0907.2189;\\
A.~Mironov and A.~Morozov, Nucl.Phys. {\bf B825} (2009) 1-37, arXiv:0908.2569;
Phys.Lett. {\bf B682} (2009) 118-124, arXiv:0909.3531;\\
N.~Drukker, J.~Gomis, T.~Okuda and J.~Teschner,
JHEP {\bf 1002} (2010) 057, arXiv:0909.1105;\\
G.Bonelli and A.Tanzini,
arXiv:0909.4031;\\
L.~Alday, F.~Benini and Y.~Tachikawa, Phys.Rev.Lett. {\bf 105} (2010) 141601, arXiv:0909.4776;\\
R.~Poghossian,
JHEP {\bf 0912} (2009) 038,  arXiv:0909.3412;\\
L.~Hadasz, Z.~Jaskolski and P.~Suchanek,
arXiv:0911.2353;
arXiv:1004.1841;\\
V.~Alba and And.~Morozov,
Nucl.Phys. {\bf B840} (2010) 441-468, arXiv:0912.2535;\\
M.~Taki,
arXiv:0912.4789;
arXiv:1007.2524;\\
S.~Yanagida, arXiv:1003.1049;
arXiv:1010.0528;\\
N.~Drukker, D.~Gaiotto and J.~Gomis
arXiv:1003.1112;\\
H.~Itoyama and T.~Oota,
arXiv:1003.2929;\\
C.~Kozcaz, S.~Pasquetti, F.~Passerini and N.~Wyllard,
arXiv:1008.1412;\\
H.~Itoyama, T.~Oota and N.~Yonezawa,
arXiv:1008.1861;\\
K.~Maruyoshi and F.~Yagi,
arXiv:1009.5553;\\
G.~Bonelli, K.~Maruyoshi, A.~Tanzini and F.~Yagi,
  arXiv:1011.5417;\\
A.~Mironov, A.~Morozov and Sh.~Shakirov,
arXiv:1010.1734;
arXiv:1011.3481


\bibitem{EY}
  T.~Eguchi and S.~K.~Yang,
  Mod.\ Phys.\ Lett.\  A {\bf 9} (1994) 2893
  [arXiv:hep-th/9407134].

\bibitem{KS}
A.~Klemm and P.~Sulkowski,
  Nucl.\ Phys.\  B {\bf 819} (2009) 400
  [arXiv:0810.4944 [hep-th]].

\bibitem{lomamo}
R.~Dijkgraaf and C.~Vafa, arXiv:0909.2453

\bibitem{varna}
 A.~Marshakov,
  arXiv:hep-th/0401199.

\bibitem{mig}
A.~Migdal,
  Phys.\ Rept.\  {\bf 102} (1983) 199.

\bibitem{dv}
R.~Dijkgraaf and C.~Vafa,
  Nucl.Phys. {\bf B644} (2002) 3, hep-th/0206255;
  Nucl.Phys. {\bf B644} (2002) 21, hep-th/0207106;
hep-th/0208048.

\bibitem{cm}
 L.~Chekhov and A.~Mironov,
  Phys.Lett. {\bf B552} (2003) 293, hep-th/0209085

\bibitem{wdvvbad}
 L.~Chekhov, A.~Marshakov, A.~Mironov and D.~Vasiliev,
  Phys.\ Lett.\  B {\bf 562} (2003) 323
  [arXiv:hep-th/0301071];
Proc. Steklov Inst.Math. {\bf 251} (2005) 254, hep-th/0506075.

\bibitem{KM}
V.~Kazakov and A.~Marshakov,
  J.\ Phys.\ A  {\bf 36} (2003) 3107
  [arXiv:hep-th/0211236].

\bibitem{mamo}
A.~Marshakov,
  Theor.\ Math.\ Phys.\  {\bf 147} (2006) 583
  [Teor.\ Mat.\ Fiz.\  {\bf 147} (2006) 163]
  [arXiv:hep-th/0601212];  
  Theor.\ Math.\ Phys.\  {\bf 147} (2006) 777
  [Teor.\ Mat.\ Fiz.\  {\bf 147} (2006) 399]
  [arXiv:hep-th/0601214].

\bibitem{HO} A.~Hanany and Y.~Oz,
  Nucl. Phys.   {\bf B452} (1995) 283
  [arXiv:hep-th/9505075].

\bibitem{GMMM} A.Gorsky, A.Marshakov, A.Mironov and A.Morozov,
Phys.Lett. {\bf B380} (1996) 75-80, [arXiv:hep-th/9603140].

\bibitem{MMMZam}
A.~Marshakov, A.~Mironov and A.~Morozov,
  JHEP {\bf 0911} (2009) 048
  [arXiv:0909.3338 [hep-th]].

\bibitem{GMMMO}
 A.~Gerasimov, A.~Marshakov, A.~Mironov, A.~Morozov and A.~Orlov,
  Nucl.\ Phys.\  B {\bf 357} (1991) 565.

\bibitem{ZamZam}
A.~Zamolodchikov and Al.~Zamolodchikov,
  Nucl.\ Phys.\  B {\bf 477} (1996) 577
  [arXiv:hep-th/9506136].

\bibitem{Gai}
D.~Gaiotto,
  arXiv:0904.2715 [hep-th].

\bibitem{EM}
 T.~Eguchi and K.~Maruyoshi,
  JHEP {\bf 1002} (2010) 022
  [arXiv:0911.4797 [hep-th]]; JHEP {\bf 1007} (2010) 081
  [arXiv:1006.0828 [hep-th]].

\bibitem{GWtk}
D.~Gaiotto, arXiv:0908.0307.

\bibitem{MMMWtk}
A.Marshakov, A.Mironov and A.Morozov,
Phys.Lett. {\bf B682} (2009) 125-129, arXiv:0909.2052.

\bibitem{Barnes}
E.~Barnes, Proc. London Math. Soc. {\bf 31} (1899) 358-381;
Phil. Trans. Roy. Soc. London {\bf A196} (1901) 265-387;
Trans. Cambridge Phil. Soc. {\bf 19} (1904) 374-425.


\bibitem{Spre}
M.~Spreafico, Journ. of Number Theory, {\bf 129} (2009) 2035.

\bibitem{Wgrav}
A.~Gerasimov, A.~Levin and A.~Marshakov,
Nucl.\ Phys.\ B {\bf 360}, 537 (1991).
\\
A.~Bilal, V.~V.~Fock and I.~Kogan,
Nucl.\ Phys.\ B {\bf 359}, 635 (1991).

\bibitem{KriW}
I.~Krichever,
Commun. Pure. Appl. Math. {\bf 47} (1992) 437
[arXiv: hep-th/9205110].

\bibitem{KMZ}
I.~Krichever, A.~Marshakov and A.~Zabrodin,
  Commun.\ Math.\ Phys.\  {\bf 259} (2005) 1
  [arXiv:hep-th/0309010].

\bibitem{qute}
L.~Faddeev and R.~Kashaev, Mod.Phys.Lett. {\bf A9} (1994) 427-434, arXiv:hep-th/9310070;\\
R.~Kashaev, arXiv:q-alg/9705021;\\
L.~Chekhov and V.~V.~Fock, Theor.Math.Phys. 120 (1999) 1245-1259,
[Teor.Mat.Fiz. 120 (1999) 511-528], arXiv:math/9908165;\\
V.~V.~Fock and A.~Goncharov, arXiv:math/0311149;\\
J.~Teschner,
  Int.\ J.\ Mod.\ Phys.\  A {\bf 19S2} (2004) 459
  [arXiv:hep-th/0303149]; arXiv:1005.2846.

\bibitem{AlZLi}
Al.~Zamolodchikov,
  Int.\ J.\ Mod.\ Phys.\  A {\bf 19S2} (2004) 510
  [arXiv:hep-th/0312279];
  arXiv:hep-th/0505063.

\bibitem{KLST}
S.~Kharchev, D.~Lebedev and M.~Semenov-Tian-Shansky,
  Commun.\ Math.\ Phys.\  {\bf 225} (2002) 573
  [arXiv:hep-th/0102180].

\bibitem{NS}
N.Nekrasov and S.Shatashvili, arXiv:0901.4748

\bibitem{KT}
K.Kozlowski and J.Teschner, arXiv:1006.2906.


\end{thebibliography}
\end{document}